\documentclass[aps,pre,twocolumn,showpacs,groupedaddress]{revtex4}
\pdfoutput=1

\usepackage{amsmath,amssymb,xspace,graphicx,xcolor}

\newcommand\gsim{\raise2.5pt\hbox{$>$}\llap{\lower2.5pt\hbox{$\sim$}}}
\newcommand\lsim{\raise2.5pt\hbox{$<$}\llap{\lower2.5pt\hbox{$\sim$}}}

\begin{document}

\title{Run-and-tumble particles with hydrodynamics:\\
sedimentation, trapping and upstream swimming}

\author{R. W. Nash$^1$,  R. Adhikari$^{1,2}$, J. Tailleur$^1$ and M. E. Cates$^1$}
\affiliation{$^1$SUPA, School of Physics, University of Edinburgh, JCMB
Kings Buildings, Edinburgh EH9 3JZ, United Kingdom\\
$^2$The Institute of Mathematical Sciences, CIT Campus, Chennai 600113, India}

\date{\today}

\begin{abstract}
  We simulate by lattice Boltzmann the nonequilibrium steady states of
  run-and-tumble particles (inspired by a minimal model of bacteria),
  interacting by far-field hydrodynamics, subject to
  confinement. Under gravity, hydrodynamic interactions barely perturb
  the steady state found without them, but for particles in a harmonic
  trap such a state is quite changed if the run length is larger than
  the confinement length: {a self-assembled pump is formed. Particles likewise confined in a narrow
channel show a generic upstream flux in Poiseuille flow: chiral swimming is not required.}
\end{abstract}

\maketitle

The motility of microorganisms raises basic physics questions that
range from local swimming mechanisms \cite{specialissue,hill,circles}
to many-body emergent phenomena \cite{viscek,pagona,loewen}. In the
latter context, even grossly simplified models represent a challenging
and active area of nonequilibrium statistical mechanics
\cite{viscek}. In some cases experimental near-counterparts to these
models can be devised in which various complicating factors (cell
division, chemotaxis, etc.)  are environmentally or genetically
suppressed \cite{berg}.

Indeed certain bacteria, including {\em E. coli}, exhibit motion which
can be idealized as a `run-and-tumble' model. Here straight `runs' at
constant speed $v$ are punctuated by sudden, rapid and complete
randomizations in direction, or `tumbles', occurring stochastically
with rate $\alpha$ \cite{berg}. The mean run length is $\ell =
v/\alpha$ and duration $1/\alpha$; at larger length and time scales
Fick's law is obeyed, with diffusivity $D=v^2/d\alpha$ in $d$
dimensions \cite{schnitzer}. This model offers an important paradigm
for a diffusion process that is fundamentally non-Brownian. Subtle
consequences of this are manifest for particles in external force
fields, such as gravity or a harmonic trap \cite{tailleur}.  In the
first case, the gravitational decay length $\lambda$ falls strictly to
zero when the gravitational force $f$ exceeds the propulsive force
$f_p$, in contrast to Brownian particles for which $\lambda = D/f$
\cite{tailleur}. In a harmonic trap ($f=-kr$), particles are strictly
confined within a radius $r^* = f_p/k$; and for $\ell \gsim r^*$ the
maximum density occurs at $r\sim r^*$ not $r=0$.  In this limit, a
particle in the trap interior rapidly swims out to $r^*$ and stays
there a long time until its next tumble \cite{tailleur}.

The qualitative physics of the aforementioned results is robust to
both a distribution in $v$, or a residual true Brownian
diffusivity. On the other hand, because there is no underlying free
energy (which would give a Boltzmann distribution as the {\em unique}
steady state), long-range hydrodynamic interactions (HI) between the
particles could have major consequences, even for steady-state
behavior. Several computational approaches to address hydrodynamics
have been developed \cite{pagona}, but none have addressed the basic
physics problems considered below: (a) sedimentation in a container
with a solid bottom; (b) confinement by a harmonic trap; and (c)
Poiseuille flow between parallel plates. These we consider at small
but finite particle density, so that in (a,b) only the {\em far-field}
hydrodynamics are important. In (c), the main hydrodynamic effect is
instead the coupling to an imposed flow.

Problems of bacteria in force fields may appear to have little direct
relevance to biology \cite{berg}. {This could explain a surprising
lack of experiments on both bacterial sedimentation, {\em e.g.}, by
centrifugation, and trapping (where the regime $\ell \gsim
r^*$ might be achieved using recent optical
methods~\cite{Lincoln}.)} We argue that these simple cases demand to
be understood before one can claim to explain more complex (and
biologically relevant) ones, and hope our work will stimulate new
experiments to help fill such gaps.

In this Letter we address run-and-tumble systems confined by gravity,
traps or walls. In the first two cases, our goal is to see whether the
nonequilibrium steady-states found without HI \cite{tailleur} survive
with HI present. For sedimentation, we find {only weak effects of HI}. In contrast, for a harmonic trap, we find that only for
$\ell \ll r^*$ is the near-Gaussian distribution seen without HI
maintained; whenever $\ell \gsim r^*$ the `density-inverted' steady
state is destroyed by HI, replaced instead by a remarkable
self-assembled pump-like structure. {We explore in detail the origin of this instability, in which the local co-alignment of swimmers causes HI to add coherently rather than with random signs, vastly enhancing their effects.} 
Thirdly, we address swimmers
confined between walls at separations $h$. {HI do not prevent swimmers
from accumulating at the walls \cite{loewen}, where a weak Poiseuille flow causes
strong upstream alignment, and hence, for $h\lesssim\ell$ a net
upstream particle flux.} This mechanism is much simpler than one
explored previously at larger $h$ \cite{hill}.

An obstacle to simulations is the expense of handling {\em near-field}
aspects of the HI \cite{pagona}. These are clearly important at high
enough density, but depend on a swimmer's precise geometry and
stroke. In contrast, the far-field flow around a swimmer, {in the
absence of any body force acting on it,} is of universal stresslet form
\cite{pagona}, with force dipole strength $s = \nu f_p a$. Here $a$ is
a hydrodynamic radius, and $\nu$ is order one; the organism is
`extensile' for $\nu >0$, `contractile' for $\nu<0$.  (Most bacteria
are extensile.) At low concentrations, modeling only the stresslet
fields should capture any universal macroscopic consequences of HI. As
shown below, this simplification drastically reduces the computational
cost.

{{\em Method:}} We use the lattice Boltzmann (LB) method to handle
the fluid momentum transport \cite{succi}, coupled to a description of
each swimmer as a pair of off-lattice point particles at fixed
separation $\nu a$ (we choose $\nu = 2$). At these points, forces
$\pm{\bf f}_p$ (of fixed magnitude, with $+$ denoting the `head' of
the swimmer) are exerted on the fluid, creating an extensile
stresslet. These forces are mapped onto the LB lattice by an optimal
discretization \cite{rupert1}, alongside any external force ${\bf f}$
which we assume to act only on the head. (This force is passed
directly onto the fluid: there is no particle inertia.)  A local fluid
velocity (excluding a self-term) is then interpolated from the lattice
to the head particle \cite{rupert1}; the swimmer co-rotates with the
local fluid and moves relative to it with an additional velocity ${\bf
u} =({\bf f}_p+{\bf f})/6\pi\eta a$. Although pointlike for forcing
purposes, our particles have finite hydrodynamic radius, $a=0.05$;
particle scale Reynolds numbers are small, as required
($5\times10^{-3}$). As shown in \cite{rupert1} our method efficiently
simulates dilute forced colloids, using $a$ values much smaller than
the lattice spacing (unity). {Indeed with $a=0.05$ we have
simulated sedimentation of $2^{16} = 65536$ swimmers at volume
fraction $\phi\sim 6.5\times 10^{-5}$, on a lattice of size $128^2\times
32$, using a serial code. In contrast, previously published works
using fully resolved algorithms are limited to at most a few hundred
particles \cite{pagona,underhill}. A fully resolved LB study of the same system
would involve $a$, and thus $L$, about 20 times
larger~\cite{rupert1}. This would require use of very large parallel
computers. With the latter, we hope in future to use our far-field
code to address situations involving millions of particles.}

{{\em Sedimentation:} We now turn to our results, first briefly
outlining those for sedimentation. (An additional figure and further
discussion are available in appendix~\ref{sec:appgrav}.) We have
computed steady state density profiles $\rho(z)$ under {a constant
gravity force $f$}, at various $w \equiv f/f_p$, in both small and
large systems (1000 and 65536 particles respectively), with hard walls
at the base and ceiling and periodic boundary conditions horizontally
in both cases. Far from a proximal region (of height $\sim \ell$ for
$w\ll 1$, in which the density is strongly perturbed by the basal
wall) our numerical density profiles closely resemble the analytic
result of \cite{tailleur} (exact in an unbounded domain without HI):
$\rho(z) \sim \exp[-z/\lambda]$ with $\lambda(w)\to 0$ as $w\to
1$. Indeed, the exponential form is maintained (within error) with HI
present, and $\lambda(w)$ has for small $w$ a very similar form to
that without HI. However as $w\to 1$, $\lambda$ seemingly remains
higher than predicted (see appendix~\ref{sec:appgrav}). Nonetheless, the
role of HI in altering the steady state remains modest: in particular,
we see no evidence of HI inducing macroscopic flow patterns (which, in
the absence of a free energy, would be possible in principle).} This
strongly contrasts with our results for traps, to which we turn next.

\begin{figure}\centerline
  {\includegraphics[width=7.5cm]{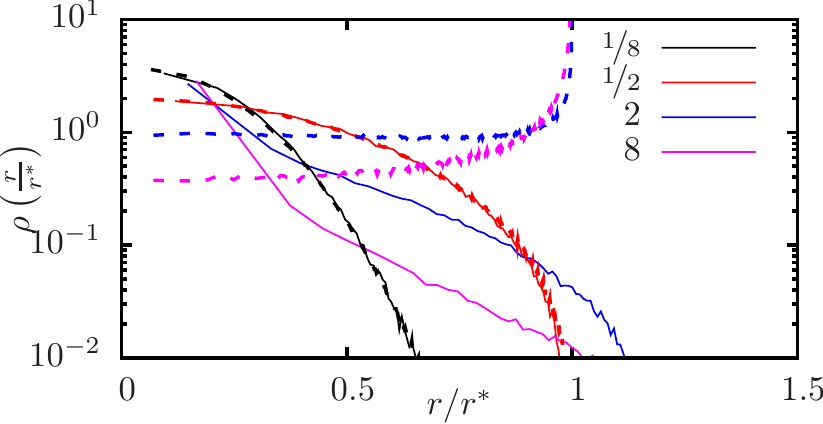}} \caption{(Color
    online) Steady state density for particles confined to harmonic
    traps. Solid lines: LB simulations; dashed lines: numerics of \cite{tailleur}. Key: $\zeta = \ell/r*$ values. Note that without HI particles cannot
    exceed the trap radius $r^*$.}
\label{fig:traps-density}
\end{figure}

\begin{figure*}
  \centerline {
    \includegraphics[width=17cm]{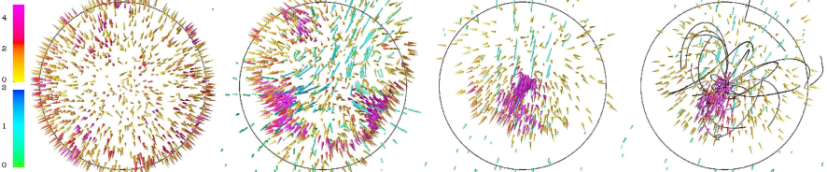}
  }
  \caption{(Color online) Simulation of $10^3$ swimmers in a harmonic
    trap. The trap radius is indicated by the circle; $\zeta =
    8$. Swimmers are marked with cones and shaded by local density in
    units of the mean (key: upper color bar). For the final image,
    randomly selected particles' trajectories over the preceding time
    interval $\Delta t = 0.5 \alpha^{-1}$ are shown.  Arrows depict
    fluid velocity colored by magnitude in units of the swim speed $v$
    (key: lower color bar). For a movie see supplementary material
    \cite{supplementary}.}
  \label{fig:trap-frame}
\end{figure*}

{{\em Traps:}} Fig.\ref{fig:traps-density} shows steady state densities $\rho(r)$ for
particles confined to harmonic traps with various ratios of run length
to confinement length, $\zeta = \ell/r^*$. These are compared to the
numerics of \cite{tailleur} without HI; the latter were also used as
the initial condition for our runs. For small $\zeta$, we see very
little effect of HI: both algorithms show the Gaussian distributions
expected for particles of diffusivity $D$ in a harmonic trap. For
larger $\zeta$ however, HI dramatically destabilize the
`density-inverted' state (with $\rho(r)$ maximal at $r^*$) that arises
without HI. Rather than approaching a steady state with zero
macroscopic flux (as the non-HI system does), the system moves to an
attractor in which rotational symmetry is broken by the formation of a
swarm of outward-swimming particles. To a good approximation the swarm
remains stationary at a radius $r_s<r^*$, where the propulsive force
balances the collective drag and the trapping force. The latter is
passed on to the fluid to create a macroscopic flow: one has in effect
a self-assembled pump. The resulting flow is mainly of stokeslet form,
with fluid flowing opposite to the swimming direction. (A slow
rotation of this direction, which might cancel the stokeslet term on
time averaging, is detectable, but the rotation rate vanishes as
$\alpha\to 0$.) When a swimmer in the clump tumbles, it is ejected in
a random direction. The velocity gradient of the stokeslet flow
rotates it to point upstream; confined by the trap, it eventually must
rejoin the swarm. A series of snapshots is given in
Fig.\ref{fig:trap-frame}, and a movie in \cite{supplementary}.

\begin{figure}
  \includegraphics[width=7.0cm]{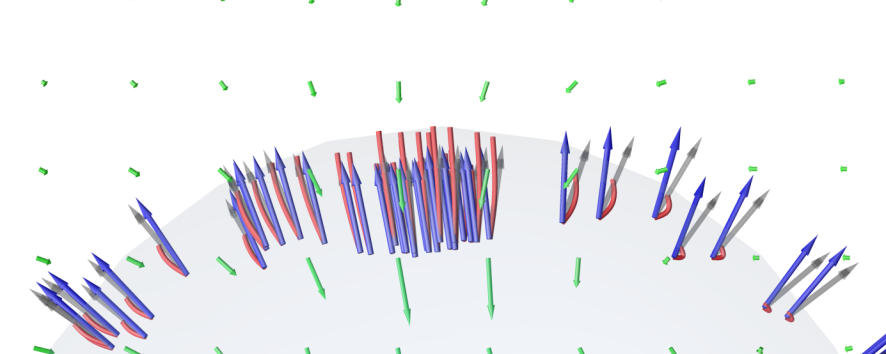}
  \caption{{(Color online) Snapshot showing the early stage instability in a simulation of 1000 swimmers, initially uniformly and randomly
  distributed on the surface of a trap. Green arrows: local flow field. Red lines: swimmer
  trajectories. Blue arrows: swimming direction. Grey arrows: radial direction. Swimmers in the dense patch (centre) move radially inwards. Those either side are initially advected away from the patch, but rotate their swimming direction towards it.}}
  \label{fig:zoominsta}
\end{figure}

{This emergent structure contains regions of high density whose
  details will depend on near-field physics that we do not resolve.
  However, the initial instability is well captured by far-field
  hydrodynamics. The end-state is likewise robust, in that {\em any} instability leading to formation of a single oriented
  swarm within the confines of a trap (harmonic or otherwise) will cause a similar
  `self-pumping' state.  To better understand the initial instability, we
  consider the limit of infrequent tumbling, $\ell/r^*\gg 1$. Here,
  without HI, the unique steady state comprises a thin uniform layer
  of outward swimmers at $r^*$, each with its propulsive force
  balanced by the external trapping force \cite{tailleur}. Coupling
  this to a solvent, we have locally a two dimensional layer of
  particles exerting inward point forces (stokeslets) on the fluid.

Because the particles on the initial shell can have not only tangential
displacements (causing density changes) but also radial and
orientational ones, a formal stability analysis of the coupled
hydrodynamics of the swimmers in this layer is not
practicable. (Moreover, noise in the run and tumble dynamics could
alter the conclusions.) Nonetheless, by a careful series of
simulations (detailed, with additional figures, in
appendix~\ref{sec:appinst}) we have identified that the instability mode is
caused initially by tangential density fluctuations on the surface of
the trap ($r=r^*$). Such density fluctuations are inevitable if the
total number of swimmers is not infinite, and lead to instability via
the following mechanism. Any surface patch that happens to be denser
than the surrounding ones will generate locally an excess
stokeslet-like flow (not cancelled by the contributions from distant
parts of the shell). This flow has two effects. First, it advects the
swimmers in the dense patch towards the center of the trap; second, it
rotates neighboring particles so that these start swimming {\it
towards} the dense patch, creating positive feedback
(Fig. \ref{fig:zoominsta}). The resulting clump creates a macroscopic
flow that sweeps the remaining particles towards itself
(Fig.~\ref{fig:trap-frame}).

On the other hand, if in our simulations we radially displace a patch
without altering its density, the feedback is negative; this also
applies for a patch in which only the orientation of the swimmers is
perturbed. Thus, although radial and orientational modes are strongly
excited in the subsequent dynamics, they are not the cause of the
initial instability (see appendix~\ref{sec:appinst}). Notably the same
instability is present for contractile rather than extensile swimmers
\cite{thesis}. This is fully consistent with the mechanism above, in
which the stokeslet from the confining force is transferred to the
fluid by a stationary swimmer (the sign of whose stresslet is then
less important).

Since the densest initial patch usually outstrips its competitors to
create a dominant stokeslet, there is little excitation of modes above
the first spherical harmonic. (This can also be explained by expanding
in such harmonics the flow arising from a nonuniform shell of point
forces \cite{thesis}.) The instability reported here is somewhat
related to that found by Crowley, who showed that dense regions in a
sedimenting suspension fall more rapidly \cite{crowley}, with the
fastest growth at long wavelengths. However, the initial force balance
is different enough to prevent a direct mapping from one instability
to the other.}

{{\em Upstream swimming:}} We turn finally to run-and-tumble particles confined between parallel
plates at separation $h$. In the absence of HI, we expect the majority
of particles to be near the walls whenever $\ell\gsim h$. {The reason
is the same as for the spherical trap, but the HI-induced instability
of the layer found in that case is suppressed here by the no-slip
boundary condition at the wall.}

The dynamics of a swimmer adjacent to a wall can be complex
\cite{berg,hill,circles}. Instead of resolving the near-field HI we
apply a truncated Lennard Jones (LJ) potential (range $0.25$ lattice
sites) which balances the normal component of the propulsive
force. The tangential component still leads to motion. In practice,
near-field HI could reduce or enhance this motion and also can, for
chiral swim-strokes, lead to circular orbits \cite{circles}. Treating
these complexities would introduce nonuniversal parameters into the
model; by neglecting them, we can clarify whether or not they are
essential in the context of Poiseuille flow.
{Here upstream swimming has been observed experimentally, and a
detailed mechanism proposed that requires chiral swimming
\cite{hill}.} Without questioning this result for the regimes
addressed in \cite{hill}, we {note} that for narrow micro-channels
($\ell \gsim h$), a much simpler mechanism is also at work. Recall
that in this regime, any swimmer in the bulk of the channel moves to
the wall layer and stays there a long time before tumbling. {On
approach to the wall any such swimmer is rotated by the velocity
gradient so as to align its swimming direction upstream.} This effect
is strong whenever the product of the wall shear rate and the transit
time across the gap is large; this equates to $u/v \gsim 1$ with $u$
the flow speed at the mid-plane.
This condition ensures that for a uniform $\rho$ ($\ell \ll h$) the
upstream wall flux is outweighed by a downstream bulk flux. However
for $\ell \gsim h$, $\rho$ is peaked at the walls, and a net upstream
flux results. {For typical particle trajectories see appendix
\ref{sec:appflow}.} Fig.\ref{fig:4} shows the region in parameter
space where there is a net upstream flux, and plots of the particle
density and mean orientation in this region. {These results are
slightly shifted quantitatively, but not qualitatively, by increasing
the LB resolution, or adopting slightly different (but still inelastic
\cite{tailleur}) wall collision rules.}

\begin{figure}
  \includegraphics{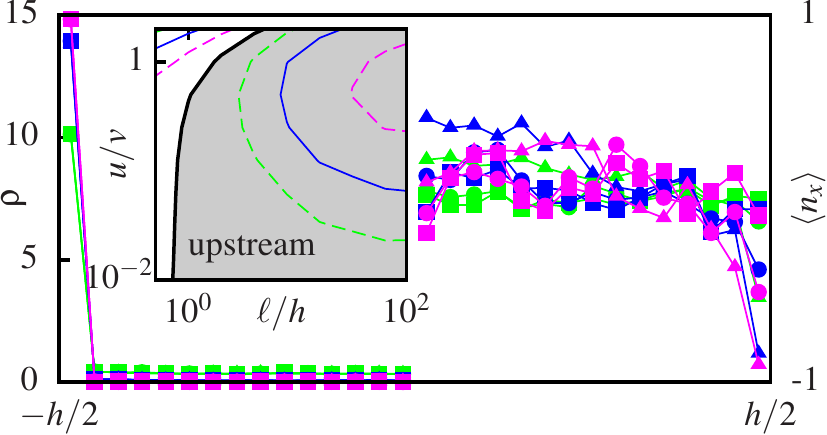}
  \caption{(Color online) Main figure: density (left) and orientation
(right) profiles.  Parameters: $u/v=1/2,1/16, 1/128$ (triangles,
circles, squares) and $\ell/h = 2,16,128$ (green, blue, pink).  Inset:
state diagram showing contours of the average swimmer speed, at
interval $v/4$. Starting from $+0.5v$ in the top left (solid blue),
the value decreases to zero (thick black), further reducing to $-0.75
v$ on the right (dashed pink). At the wall, particles feel a truncated
LJ potential of range 0.25.}
  \label{fig:4}
\end{figure}

{{\em Conclusion:}} Above we have presented results from an efficient
hydrodynamic simulation of dilute run-and-tumble swimmers. Under
gravity we find only perturbative effects of HI on the flux-free
steady state of \cite{tailleur}. Much stronger effects are found in
geometries where, without HI, all particles are swimming locally in
the same direction. {This result is consistent with a crude power
counting argument: for a sheet of swimmers, the $1/r^2$ velocity
contributions from each stresslet cannot give a large effect unless
they add coherently. This happens for the trap and the parallel plate
geometry, but only when the run-length exceeds the confinement
length.} For harmonic traps, the flux-free steady state is {then}
replaced by one in which a symmetry-breaking swarm of swimmers acts as
a pump.  For swimmers in microchannels, outward-oriented particles at
the confining walls are aligned by shear to create an upstream
particle flux without the need for chiral motion. We hope that these
predictions will stimulate new quantitative experiments on the
fundamental physics of self-propelled micro-organisms in suspension.

We thank D. Marenduzzo, A. Morozov, K. Stratford and W. Poon for
discussions and EPSRC EP/E030173 and EP/H027254 for funding. MEC is
funded by the Royal Society.

\appendix{}

\section{Results for sedimentation}
\label{sec:appgrav}
\begin{figure}
  \centerline {
    \includegraphics[width=7.5cm]{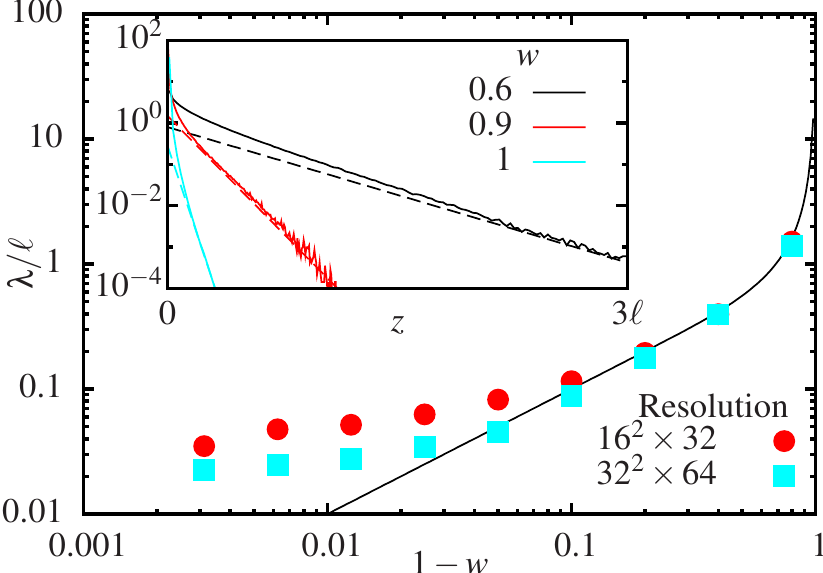}
  } 
  \caption{(Color online) Main figure: $\lambda/\ell$ vs. $1-w$ from
    fits to the distal part of $\rho(z)$ in a system of 1000
    particles. Solid line: prediction of \cite{tailleur}; circles,
    squares: simulations with HI at two resolutions. Inset: profiles
    of the swimmer density with the fitted exponentials (dashed). For
    better statistics, we repeated the runs for $w = 0.9,0.95,0.975$
    in a $128^2\times 32$ system with $2^{16}=65536$ swimmers and
    found extremely similar $\lambda$ values.  We checked that for
    $w\ge 1$ the profile collapses entirely if we switch off HI after
    its formation.} \label{fig:sedi} \end{figure}

Fig.\ref{fig:sedi} shows results for steady state density profiles
$\rho(z)$ under {a constant gravity force $f$}, at various $w \equiv
f/f_p$, in a system of 1000 particles with hard walls at the base and
ceiling and periodic boundary conditions horizontally.  As discussed
in the main text, far from a proximal region the exponential form
predicted without HI in \cite{tailleur} is maintained (within error)
with HI present. However as $w\to 1$, $\lambda$ seemingly remains
higher than predicted. The details depend however on the resolution
used (Fig.\ref{fig:sedi}) and our studies suggest that the larger
fitted $\lambda$ is in fact a proximal effect caused by the random
stirring of the fluid by particles near the wall, causing upward
advection of some of the particles above. If so, we expect this effect
to die out at larger distances with the asymptotic decay length
$\lambda$ reverting to its non-HI value.  In any case, the role of HI
in altering the steady state appears modest: in particular, we see no
evidence of HI inducing macroscopic flow patterns (which, in the
absence of a free energy, would be possible).

\section{The origin of the trap instability}

\label{sec:appinst}
\begin{figure}
  \includegraphics[width=.9\columnwidth]{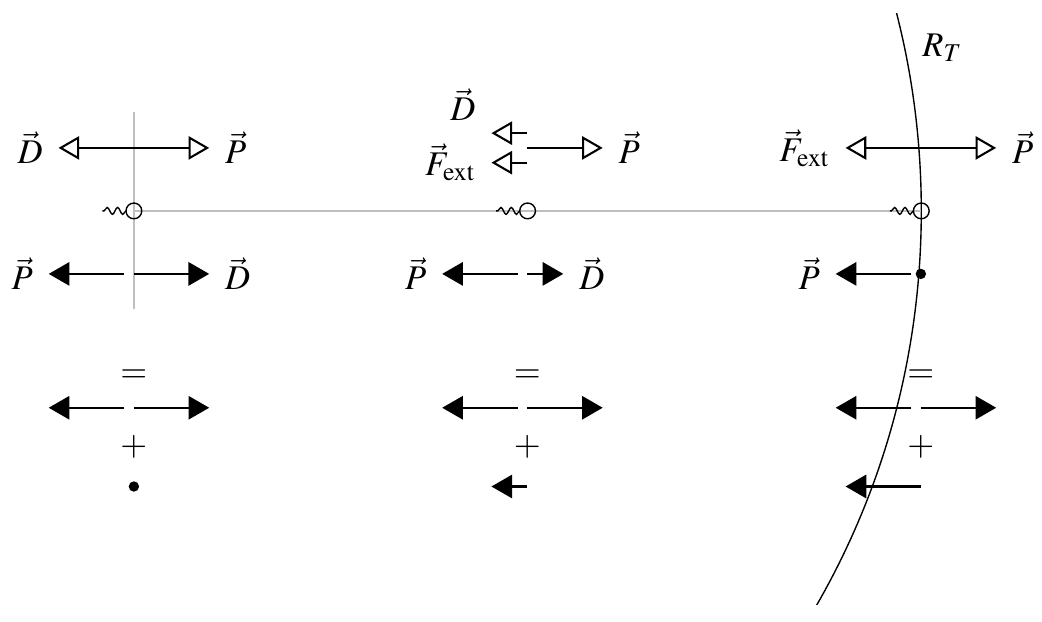}
  \caption{Forces on a swimmer as function of its distance to the
    center of the trap. The upper, open-headed arrows show the forces
    on the swimmer's body and the lower, filled arrows show the forces
    on the fluid. $P$ is the propulsive force, $D$ the drag and
    $F_{\rm ext}$ the external, trapping force. The very bottom arrows
    decompose the swimmer's effect on the fluid into stokeslet and
    stresslet components, illustrating that the external, trapping
    force is effectively passed onto the fluid.}
\label{fig:effectofswimmer}
\end{figure}

In the following we show how the disorder in the position of swimmers
on a shell generates the instability, and compare it with other
possible mechanisms that are shown to be less relevant. To do so we
consider an initial condition where the swimmers have swum as far as
they can and are thus located on the surface of a sphere of radius
$r^*=f_p/k$. In the absence of any flow, the only forces acting on the
swimmers are their propulsive forces and the trapping force. Because
both the fluid and the swimmers are at rest, there is no drag force
and the only force felt by the fluid is the propulsive force:
effectively the external trapping force has been passed onto the fluid
and the effect of the swimmer on the fluid {is to create a static,
inward pointing stokeslet whose amplitude equals the propulsive force}
(see Fig.~\ref{fig:effectofswimmer}).

\begin{figure}
  \includegraphics{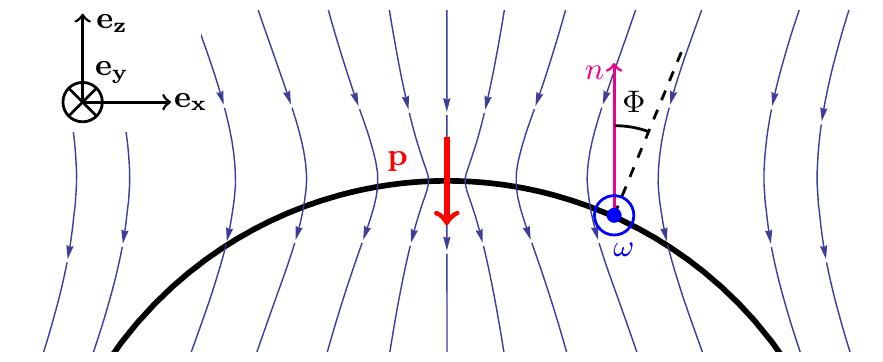}
  \caption{We consider the flow field generated by a stokeslet ${\bf
  p}$ (red arrow) held fixed at the north pole of a sphere of radius
  $r^*$ and its effect on a swimmer located at Euler angles $\theta$
  and $\varphi=0$, whose orientation (magenta arrow) makes an angle $\Phi$
  with the normal to the sphere. The vorticity is indicated in blue
  and tend to rotate the swimmer towards the stokeslet.}
  \label{schema}
\end{figure}

Because our swimmers are point-like particles at zero Reynolds {number}, the
flow created by $N$ swimmers is simply the superposition of the flow
created by each {one} of them. To understand the effect of such flows on a
swimmer, let us consider the simpler situation of an inward pointing
stokeslet of magnitude $p$ held fixed at the north pole of a sphere of
radius $r^*$ and a swimmer located at an Euler angle $\theta,\varphi=0$
on the surface of this sphere (see Fig.~\ref{schema}). The flow
field created at this point by the stokeslet is given
by~\cite{pozrikidis}
\begin{equation}
  \vec u = \frac{p}{16 \pi \eta r^* \sqrt{2 (1-\cos\theta)}} \begin{pmatrix} \sin\theta\\0\\ \cos\theta- 3\end{pmatrix}
\end{equation}
while the corresponding vorticity is:
\begin{equation}
  \vec \omega=\frac{p}{8 \sqrt{2} \pi (r^*)^2 \eta (1-\cos
    \theta)^{3/2}} \begin{pmatrix} 0\\-\sin\theta\\0 \end{pmatrix}
\end{equation}

For small $\theta$, that is when the swimmer is in the {vicinity} of
the stokeslet, the flow has two effects: the swimmer is advected
inwards ($u_z<0$) and rotated by the vorticity field {\em towards} the
stokeslet ($\omega_x=\omega_z=0$ while $\omega_y<0$, the swimmer
rotates counterclockwise on figure \ref{schema}). As time goes on, the
swimmer thus also start swimming {\em towards} the stokeslet. This
argument can be {made} more quantitative by considering the dynamics of the
swimmer ({neglecting the effect of the swimmer on the flow}):
\begin{equation}
  \dot \Phi = \omega;\quad \gamma (\dot {\bf r}-{\bf u})= {\bf f_p}+{\bf f_t}
\end{equation}
where $\gamma$ is the effective friction coefficient of the
swimmer~\cite{thesis}, ${\bf f_p}$ the propulsive force of the swimmer
and ${\bf f_t}$ the trapping force. Let us then introduce ${\bf
e_n}=\sin\theta {\bf e_x}+\cos\theta {\bf e_z}$ and ${\bf
e_\theta}=\cos\theta {\bf e_x}-\sin\theta {\bf e_z}$ the unit
vectors normal and tangent to the sphere at the position of the
swimmer, so that the position of the swimmer is ${\bf r}=r {\bf
e_n}$. Using the expression of the flow, the equations of motion of
the swimmer can be rewritten
\begin{equation}
  \begin{aligned}
    \label{eqn:earlydynamics}
    \dot r &= \frac{p} {16 \pi \eta R \sqrt{2 (1-\cos\theta)}} (1-3\cos\theta)-f \gamma^{-1} (1-\cos\Phi)\\
    r \dot\theta &= \frac{3 p \sin\theta} {16 \pi \eta R \sqrt{2 (1-\cos\theta)}}-\gamma^{-1} f \sin\Phi\\
    \dot\Phi &= \frac{p \sin\theta}{16 \sqrt 2 \pi \eta R^2  (1-\cos\theta)^{3/2}}
  \end{aligned}
\end{equation}

Let us now consider the dynamics of a swimmer that is initially
pointing out radially ($\Phi=0$) and is close to the stokeslet, say
$\theta \lsim \pi/ 4$. Initially, one finds as expected $\dot r <0$ so
that the swimmers falls inwards. The second equation show that,
initially, the flow advects the swimmer {\em away} from the stokeslet
($\dot \theta\geq 0$ at $t=0$). As time goes on however, the angle
$\Phi$ increases (third equation) and $\dot \theta$ will change
sign. At larger time, the swimmer is indeed {\em swimming} towards the
stokeslet. When the particles move away from the surface of the trap,
equations (\ref{eqn:earlydynamics}) become more complex and we do not
attempt an analytic approach and include these equations only to
illustrate the qualitative behaviour of the swimmers.
\begin{figure}[h]
  \includegraphics[width=\columnwidth]{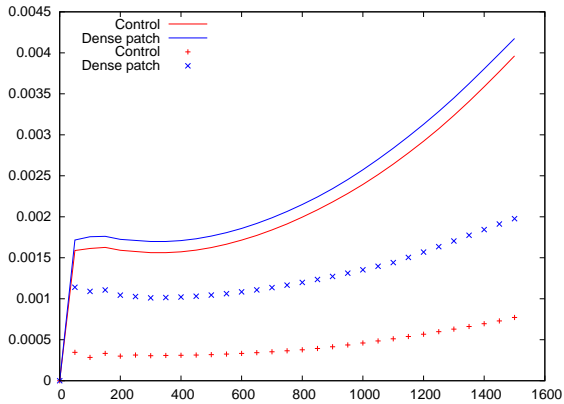}
  \caption{Comparison of the flow fields generated by a control
  simulation, starting from a random uniform initial condition, and one 
  with the densest patch on the top. The solid lines correspond to
  $\langle|v|_\infty\rangle$ and are identical for the two sets of
  simulations. This is expected since the initial condition with the
  densest patch on the top aims at being a simple rotation of the
  uniform one. The discrete symbols correspond to
  $|\langle v \rangle|_\infty$. In this case, the control decreases much
  more that the simulation with the densest patch on the top which
  shows that density fluctuation on a scale attained by the random
  distribution of swimmers suffice to control the instability. }
  \label{fig:densevscontrol}
\end{figure}
\begin{figure*}[t]
  \includegraphics[width=\textwidth]{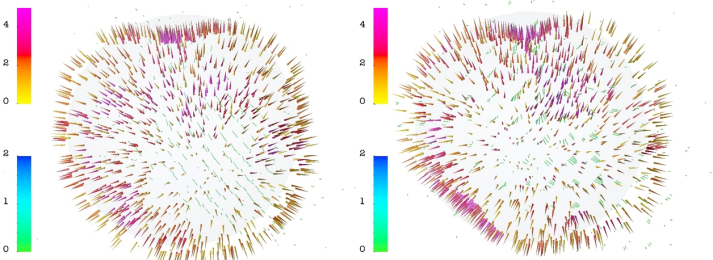}
  \caption{Comparison between a random uniform initial condition
  (left) and one where the densest patch is {`seeded'} on the top
  (right). The random uniform one has been rotated so that the
  instability starts on the top. We see very similar flow patterns and
  displacement of swimmers close to the north pole. Note that because
  of secondary dense patches, the rest of the configuration can
  slightly differ.}
  \label{fig:directcomp}
\end{figure*}

Let us now come back to the problem of $N$ swimmers in the absence of
any external flow. Firstly, note that {by symmetry one sees} that a
perfectly uniform layer of stokeslets cannot create any flow and is
thus in mechanical equilibrium. Because we consider a finite number of
swimmers, the density of stokeslets on the surface on the sphere
cannot, however, be uniform but can be decomposed as the sum of two
contributions: a perfectly uniform layer of inward pointing
stokeslets, representing the {\it average} density, and a layer of
inward and outward pointing stokeslets representing the more dense and
less dense regions, i.e., the density {\em fluctuations}. Close to the
denser regions, one can thus expect to see the scenario described
above for the motion of a swimmer in an inward pointing stokeslet
flow. This is indeed what we see in a simulation of 1000 swimmers
randomly distributed on the boundary of the trap (see Fig.~3). Around
the denser patch located at the top, the {far-field flow created by
the swimmers resembles that of a stokeslet.}\footnote{because the
dense patch has a finite size the {near field is} however slightly
different.} Furthermore, the trajectories of the neighboring swimmers
resemble the one described above for a single swimmer in a stokeslet
flow. Density fluctuations within the surface of the trap are thus a
strong candidate for the origin of the instability.

{To confirm this role} we compare simulations where the densest patch
of the initial condition is placed at the top of the trap with those
obtained from random uniform initial conditions. Let us first specify
what we mean by a dense patch. Since there are 1000 swimmers, every
patch of surface $S$ contains on average $N_s=250 S/[\pi (r^*)^2]$
swimmers, with fluctuations of the order of $\sqrt{N_s}$. {We thus
choose a patch size $S$ that is a small fraction of the sphere but
large enough that fluctuations are not much larger than the mean
density}. We achieve this by considering a spherical cap defined by
$\theta\leq \pi/12$, which contains on average $17$ swimmers. To cover
the sphere with patches of this size, one would need roughly 60
patches, which means that the densest patch contains on average 27
swimmers.~\footnote{This can be checked by negelecting the
correlations between the number of swimmers in the different patches
coming from the total constrains, and using cumulative generating
function of the binomial distribution.} {Our `dense patch' initial
condition is thus constructed as follows: each swimmer is placed
uniformly on the sphere with probability $p$ or uniformly on the top
spherical cap with probability $1-p$. In practice, $p$ is chosen so
that the average number of swimmers in the spherical cap is $27$. We
then want to compare the instability between this initial condition
and the one where swimmers are uniformly randomly distributed (from
now on, we refer to this initial configuration as `control').  If
tangential density fluctuations are not the cause of the instability,
the statistical outcome of the `dense patch' and `control' simulations
should be the same.}

To compare the instability, we can look at the generated flow fields.
From our lattice Boltzmann simulations we get a 3D grid with fluid
velocities ${\bf v}_i$ at the nodes. Any measure of the intensity of
the flow yields qualitatively similar results and we thus consider
$|{\bf v}|_{\infty}\equiv \sup_i |{\bf v}_i|$. In figure
\ref{fig:densevscontrol} we report $\langle |{\bf v}|_\infty\rangle$
and $|\langle {\bf v} \rangle|_\infty$, where the averages are done
over 50 runs both for the control and the dense patch cases. We see
that the two initial conditions give very similar $\langle |{\bf
  v}|_\infty\rangle$: this validates our `dense patch' initial
condition which is not very different, once rotated appropriately,
from a `control' one. Since the random uniform case is perfectly
isotropic, we expect the flow field to vanish when averaged over many
runs.  When comparing $|\langle {\bf v} \rangle|_\infty$, we indeed
see that such is the case for the control; by increasing further the
number of runs over which the average is done, the non-zero residual
flow would continue to decrease. For the dense patch case however, the
flow remains much larger.  Because the denser patch at the top has to
compete with other dense patches, there is some randomness in the
orientation of the flow generated by the instability and the intensity
of the flow has decreased. It remains however three times larger than
the control case, which shows that fluctuations of density of the scale
of those observed in the simulations are sufficient to {create} the
instability.  

A less quantitative but more direct comparison can be
done by starting a control simulation, locating the patch where the
instability starts and rotating the view in such a way that it is at
the top. One can then compare it with a `dense patch' run (see figure
\ref{fig:directcomp}). {Both configuration are indeed very similar, confirming the role of tangential density fluctuations in creating the instability.}

\begin{figure}[h]
  \includegraphics[width=\columnwidth]{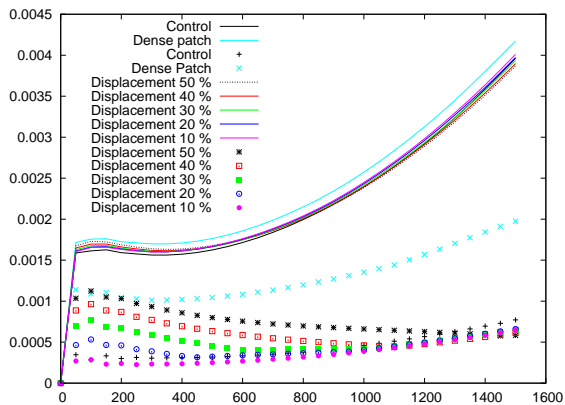}
  \caption{Comparison between control, dense patch, and {radially}
    displaced initial conditions for displacements ranging from 10\%
    to 50\% of the trap radius.  Solid lines correspond to $\langle |
    {\bf v} |_\infty \rangle$ whereas symbols correspond to $ |\langle
    {\bf v} \rangle |_\infty$. {All initial conditions produce
    similar flow (solid line), and thus similar instability.}  The
    averaged flows of the displaced initial conditions and the control
    are the same at large times, showing that these perturbations have
    a much smaller effect on the setup than the one due to density
    fluctuations. Note {that the non-zero initial flow is caused
    by particles swimming back to the border of the trap.}}
\label{fig:displaced}
\end{figure}

{
\subsection{Other type of perturbations} 

In addition to the non-uniformity of the density of swimmers
on the surface of the trap, other sources of perturbation could be
responsible for the instability. First, the swimmers can fluctuate in radial position (`radial disorder'), and
second they may not all be strictly pointing outwards (`angular
disorder'). We now show below that these fluctuations are less relevant than the tangential density fluctuations.

\subsubsection{Radial Displacement}

To study the role of radial displacement, we place the swimmers at
random over the whole sphere uniformly and then move downwards those
in the same spherical cap as previously contained the dense set of
bacteria. (There is no longer an excess density in this cap, however.) Again} we compare the instability by looking at the flow
fields. As can be seen in figure \ref{fig:displaced}, $\langle | {\bf v}
|_\infty \rangle$ is left unchanged for displacements up to half the
radius of the trap. This means that displacements of swimmers, even
much larger those spontaneously observed in simulations, do not seem
to generate any kind of instability with a stronger effect than the
one seen in their absence. Let us now turn to the comparision of
$|\langle {\bf v} \rangle |_\infty$ to look for systematic effects. At large
times, the flow field is the same as the control one and much smaller
than that of the dense patch; this perturbation does not {fix} the
direction of the instability. Hence radial disorder, even for large
displacements, does not control the instability observed from the
random initial configurations.

\begin{figure}
  \includegraphics[width=\columnwidth]{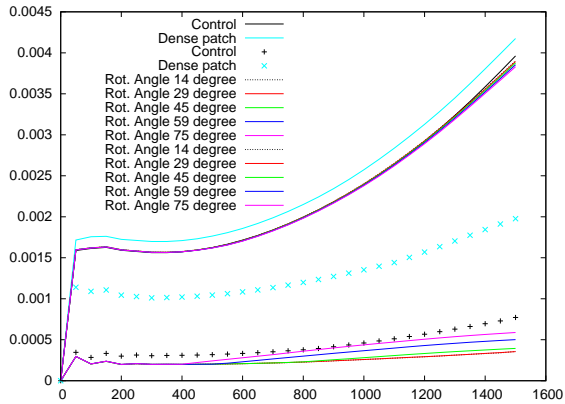}
  \caption{Comparison between control, dense patch and randomly
    rotated initial conditions for several rotation angles. Solid
    lines correspond to $\langle | {\bf v} |_\infty \rangle$ whereas
    symbols correspond to $ |\langle {\bf v} \rangle |_\infty$. All
    initial conditions produces similar strength of flow, and thus of
    instability. The average flow of the random rotations equals that
    of the control, showing that these perturbation have a much
    smaller effect on the setup than the density fluctuations.}
  \label{fig:rotationr}
\end{figure}

\subsubsection{Rotation}\

\begin{figure}
  \includegraphics[width=\columnwidth]{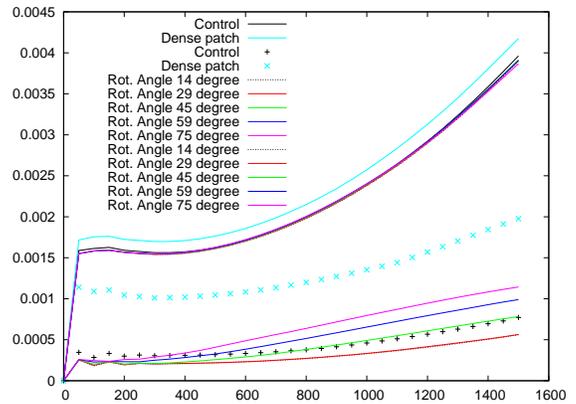}
  \caption{Comparison between control, dense patch, and uniformly
    rotated initial conditions for several rotation angles. Solid
    lines correspond to $\langle | {\bf v} |_\infty \rangle$ whereas
    symbols correspond to $ |\langle {\bf v} \rangle |_\infty$. All
    initial conditions produce similar strength of flow, and thus of
    instability. The average flow of the uniformly rotated initial
    condition is however of the same order as the control one, showing
    that these perturbation have a much smaller effect on the setup
    than the density fluctuations. }
  \label{fig:rotationu}
\end{figure}

{To investigate the effect of disorder in the swimmers
orientation, we first place the swimmers at random over the
whole surface uniformly and then modify the orientation of the
swimmers that are in the top spherical cap defined by $\theta\leq
\pi/12$.} We either generate randomly a rotation matrix and apply it to all
these simmwers (uniform rotation) or we pick up one rotation matrix
per swimmer (random rotation). The strength of the perturbation is
controlled by varying the average rotation angle of the distribution
of {the} rotation matrix. {In both cases, $\langle| {\bf v}|_\infty\rangle$
perfectly overlaps with the control (see figure \ref{fig:rotationr}
and \ref{fig:rotationu}), while the quantity $|\langle
{\bf v}\rangle|_\infty$ is as small as the control. These results confirm that the initial rotation has no part to play in triggering the instability.}

\if{
All this indicates that the relevant mechanism for the instability is
the density fluctuations on the surface of the trap. They should
generate swimmer trajectories similar to those observed numerically,
they are large enough to control the direction of the instability and
they dominate all the other types of instability that we have
considered.
}\fi

\begin{figure}[h]
  \centerline {
    \includegraphics[width=7.0cm]{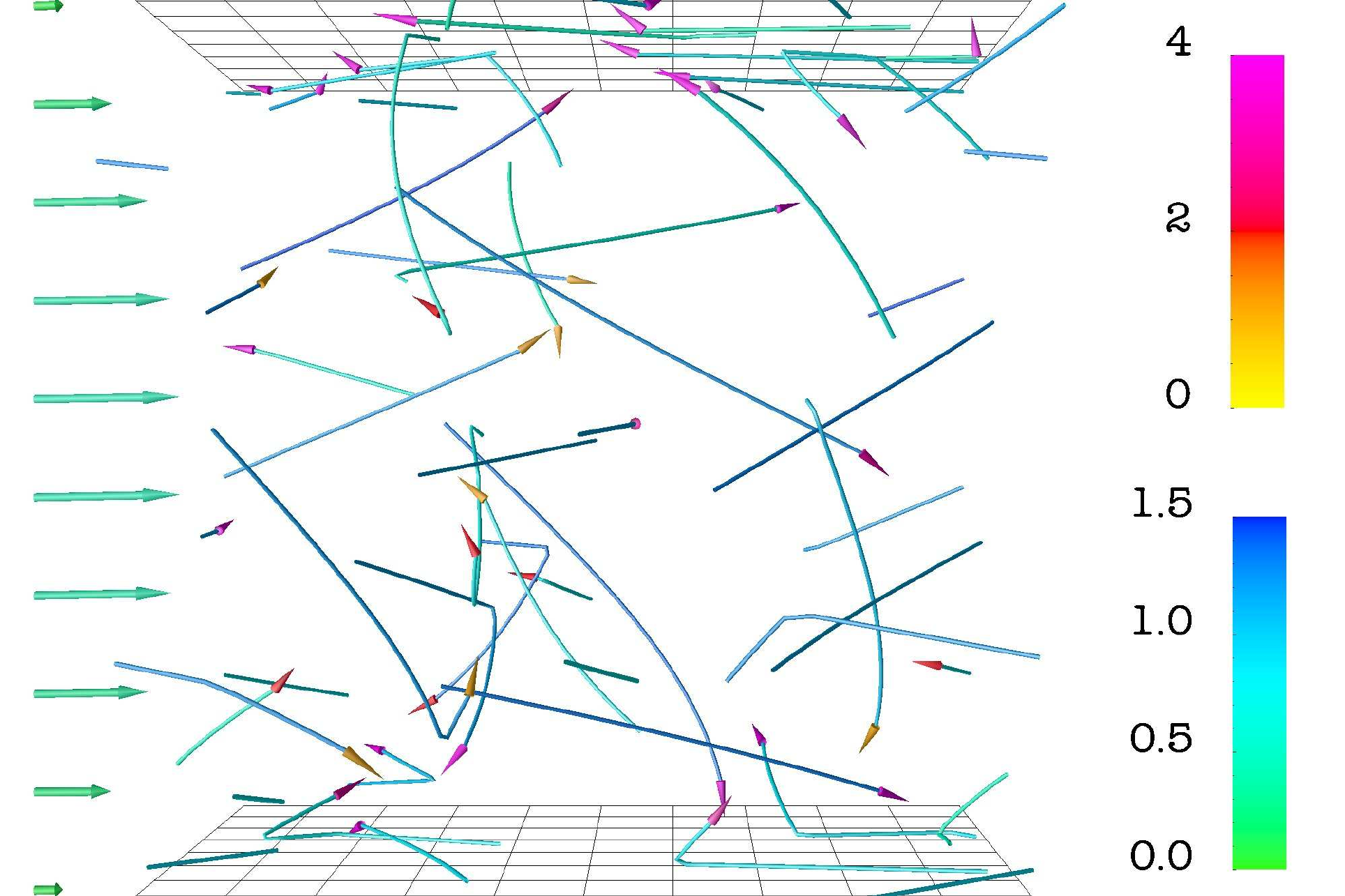}
  }
  \caption{(Color online) 40 randomly chosen swimmers (of 1000
    simulated) in a Poiseuille flow between plates. Swimmers are
    marked with cones and shaded by relative local density (key: upper
    color bar). Trajectories over the preceding time interval $\Delta
    t = 0.5 \alpha^{-1}$ are shown, colored by magnitude of lab-frame
    speed in units of $v$ (key: lower color bar). The flow profile is
    shown on the left (same key).}
  \label{fig:5}
\end{figure}

\section{Brownian Motion}
Interestingly, our stability analysis allows us to evaluate the impact
of Brownian motion on the trap instability. For {\it E. coli}, thermal
diffusivity is $D\simeq 0.3\, \mu{\rm m}^2/{\rm s}$ whereas its
rotational counterpart is $D_{\rm rot} \simeq 0.15\, {\rm rad}^2/{\rm
s}$~\cite{SFBK1998}. Over the time scale of a second, which is the
average time between two tumbles, this corresponds to diffusion length
and angle of about half a micrometer and 20 degrees. The simulations
presented in the previous subsections shows that these perturbations
are negligible when compared to the density fluctuations. Brownian
motion should thus not affect the trap instability.

\section{Channel Flow}
\label{sec:appflow}
An additional figure for channel flow, showing swimmer trajectories, is given as Fig.~\ref{fig:5}.


\begin{thebibliography}{99}

\bibitem{specialissue}  R. Golestanian and A. Ajdari, Phys. Rev. Lett. {\bf 100}, 038101 (2008); P. Garstecki and M. Cieplak, J. Phys. Cond. Mat. {\bf 21}, 200301 (2009) and references therein.
\bibitem{hill} J. Hill et al, Phys. Rev. Lett. {\bf 98}, 068101 (2007)
\bibitem{circles} W. R. DiLuzio et al., Nature {\bf 435}, 1271 (2005); S. van Teeffelen and H. L\"owen, Phys. Rev. E {\bf 78}, 020101(R) (2008).
\bibitem{viscek} A. Czirok, A.-L. Barabasi and T. Vicsek, Phys. Rev. Lett. {\bf 82}, 209-212 (1999); J. Toner, Y. H. Tu and S. Ramaswamy, Annals of Phys. {\bf 318}, 170-224 (2005); A. Sokolov et al., Phys. Rev. Lett. {\bf 98}, 158102 (2007).
\bibitem{pagona}I. Llopis and I. Pagonabarraga, EPL {\bf 75}, 999 (2006);
C. M. Pooley, G. P. Alexander and J. M. Yeomans, Phys. Rev. Lett. {\bf 99}, 228103 (2007) 
T. Ishikawa and T. J. Pedley, Phys. Rev. Lett. {\bf 100}, 088103 (2008)
\bibitem{underhill} Our work is similar in spirit to that of
P. T. Underhill, J. P. Hernandez-Ortiz and M. D. Graham, Phys. Rev. Lett. {\bf 100}, 248101 (2008), and D. Saintillan and M. J. Shelley, Phys. Rev. Lett. {\bf 99}, 058102 (2007). These both use far-field Green function methods and address several thousand particles. However, we expect our LB-based approach to prove more adaptable in complex geometries and more suited to parallelization \cite{succi}.

\bibitem{loewen} H. H. Wensink and H. L\"owen, Phys. Rev. E {\bf 78}, 031409 (2008).
\bibitem{berg}  H. C. Berg, {\em E. coli in Motion}, Springer, NY 2004.
\bibitem{schnitzer} M. J. Schnitzer, Phys. Rev. E {\bf 48}, 2553-2568 (1993).
\bibitem{tailleur} J. Tailleur and M. E. Cates, Phys. Rev. Lett. {\bf 100}, 218103 (2008); EPL {\bf 86}, 60002 (2009).
\bibitem{Lincoln} B. Lincoln {\em et al.}, Biomed. Microdevices {\bf 9} 703-710 (2007)
\bibitem{succi} S. Succi, {\em The Lattice Boltzmann Equation for Fluid Dynamics and Beyond}, Clarendon, Oxford (2001).
\bibitem{rupert1} R. W. Nash, R. Adhikari and M. E. Cates, Phys. Rev. E {\bf 77}, 026709 (2008).
\bibitem{supplementary} {See supplementary material for movies
http://prl.aps.org/supplemental/PRL/v104/i25/e258101}
\bibitem{thesis} R. Nash, PhD Thesis, University of Edinburgh (2009).
\bibitem{crowley} {J. M. Crowley, Phys. Fluids {\bf 19}, 1296 (1976);
R. Lahiri and S. Ramaswamy, Phys. Rev. Lett. {\bf 79}, 1150 (1997).}
\bibitem{pozrikidis} C. Pozrikidis {\it Introduction to theoretical and computational fluid dynamics}, (Oxford University Press, Oxford) 1997
\bibitem{SFBK1998} S.P. Strong, B. Freedman, W. Bialek, R. Koberle, Phys. Rev. E {\bf 57}, 4604 (1998)
\end{thebibliography}
\end{document}